\documentclass[12pt]{article}

\begin{document}

\newcommand{\be}{\begin{equation}}
\newcommand{\ee}{\end{equation}}
\newcommand{\bea}{\begin{eqnarray}}
\newcommand{\eea}{\end{eqnarray}}
\newcommand{\ack}[1]{[{\bf Pfft!: #1}]}

\newcommand{\eref}[1]{eq.\ (\ref{eq:#1})}
\def\NPB{{\it Nucl. Phys. }{\bf B}}
\def\PL{{\it Phys. Lett. }}
\def\PRL{{\it Phys. Rev. Lett. }}
\def\PRD{{\it Phys. Rev. }{\bf D}}
\def\CQG{{\it Class. Quantum Grav. }}
\def\JMP{{\it J. Math. Phys. }}
\def\SJNP{{\it Sov. J. Nucl. Phys. }}
\def\SPJ{{\it Sov. Phys. J. }}
\def\JETPL{{\it JETP Lett. }}
\def\TMP{{\it Theor. Math. Phys. }}
\def\IJMPA{{\it Int. J. Mod. Phys. }{\bf A}}
\def\MPL{{\it Mod. Phys. Lett. }}
\def\CMP{{\it Commun. Math. Phys. }}
\def\AP{{\it Ann. Phys. }}
\def\PR{{\it Phys. Rep. }}

\hyphenation{Min-kow-ski}
\hyphenation{cosmo-logical}
\hyphenation{holo-graphy}
\hyphenation{super-symmetry}
\hyphenation{super-symmetric}

\rightline{gr-qc/0305072}
\rightline{VPI-IPPAP-03-07}
\rightline{CERN-TH/2003-109}
\centerline{\Large \bf }\vskip0.25cm
\centerline{\Large \bf }\vskip0.25cm
\centerline{\Large \bf Deconstructing the Cosmological Constant}\vskip0.25cm
\vskip 1cm

\renewcommand{\thefootnote}{\fnsymbol{footnote}}
\centerline{{\bf Vishnu Jejjala,${}^{1}$\footnote{vishnu@vt.edu}
Robert G. Leigh,${}^{2,3}$\footnote{rgleigh@uiuc.edu} 
and
Djordje Minic${}^{1}$\footnote{dminic@vt.edu}
}}
\vskip .5cm
\centerline{${}^1$\it Institute for Particle Physics and Astrophysics}
\centerline{\it Physics Department, Virginia Tech}
\centerline{\it Blacksburg, VA 24061, U.S.A.}
\vskip .5cm
\centerline{${}^2$\it CERN-Theory Division}
\centerline{\it CH-1211, Geneva 23, Switzerland}
\vskip .5cm
\centerline{${}^3$\it Department of Physics}
\centerline{\it University of Illinois at Urbana-Champaign}
\centerline{\it 1110 W. Green Street, Urbana, IL 61801, U.S.A.}

\vskip .5cm

\begin{abstract}
Deconstruction provides a novel way of dealing with the notoriously
difficult ultraviolet problems of four-dimensional gravity.
This approach also naturally leads to a new perspective on the holographic
principle, tying it to the fundamental requirements of unitarity and
diffeomorphism invariance, as well as to a new viewpoint on the cosmological
constant problem.
The numerical smallness of the cosmological constant is implied by a unique
combination of holography and supersymmetry, opening a new window into the
fundamental physics of the vacuum.
\end{abstract}

\setcounter{footnote}{0}
\renewcommand{\thefootnote}{\arabic{footnote}}

\newpage

The validity of general relativity as a {\it classical} theory, at least
at reasonable length scales is by now beyond any doubt, yet a completely
satisfying {\it quantum} theory of gravitation remains elusive.
The difficulty may be understood on many levels.
The most straightforward approach, that of treating general relativity
as a local four-dimensional field theory and quantizing it as such,
fails unequivocally.
The gravitational coupling, $G_N$, is a dimensionful quantity that renders
the short-distance structure of the theory meaningless.
Thus, at best, general relativity should be regarded as a four-dimensional
{\it effective} field theory that is replaced by something else at short
distances, for example, a well-defined {\it perturbative} quantum theory
of gravity, such as string theory.

Yet, all is not well, even apart from the basic open question of how to
formulate a background independent non-perturbative version of quantum
gravity.
The low-energy effective field theory makes predictions wildly inconsistent
with observation.
Most notably, when coupled to matter degrees of freedom, the energy density
of the vacuum is extremely large, scaling with the largest available energy
in the theory.
This is the essence of the cosmological constant problem.
The insidiousness of the renormalization of the cosmological constant
means that it is not even sufficient to find a principle that would set
the vacuum energy to some small value at a given ultraviolet (UV) scale;
rather it must be canceled all the way into the infrared (IR).

It has recently become clear that quantum gravitational systems display
features that cannot be accommodated by local four-dimensional field
theories.
In particular, the holographic principle \cite{holo} asserts that the
degrees of freedom of such four-dimensional gravitational systems are
better accounted for by three-dimensional data.
This principle stems from the well-known non-extensive properties of the
Bekenstein-Hawking entropy \cite{bhent}
\be
{\cal S} = \frac{A}{4G_N},
\ee
which scales as the area, not the volume, of a given region of space.
Just how holography might be implemented is a matter of some debate,
but simple examples, possessing a high degree of symmetry, have been well
explored; this is what underlies the duality between gravitating systems
on anti-de Sitter (AdS) background geometries and conformal field theories
(CFT) in one fewer dimension \cite{ads}.

If holography is to be taken seriously, we should look to three-dimensional
theories for guidance.
Recent astrophysical observations of the cosmic microwave background
radiation \cite{WMAP} and distant supernovae \cite{SN} together suggest that
the expansion of the universe is accelerating and that this acceleration
is being driven by a ``dark energy,'' which comprises three quarters of
the total energy density of the universe.
The leading candidate for dark energy is the energy in the vacuum itself,
and the observed value points to a positive small cosmological constant.
An extension of the ideas underlying the dualities mentioned above would
then seem to suggest looking for a de Sitter/CFT correspondence \cite{dscft}.
It is not clear however, what three-dimensional CFT would be capable of
fully describing the present state of our Universe.

However, there is another possibility based on the idea of {\it
deconstruction} \cite{nima}.
In this framework, one imagines that the short distance regime of
a four-dimensional field theory is described by a three-dimensional
theory.
The most amazing possibility is that by introducing supersymmetry
into the {\it three}-dimensional theory, it is possible that the {\it
four}-dimensional theory has a small cosmological constant!
This statement relies on specific properties of three-dimensional
supersymmetry, first noticed by Witten \cite{wit1}.
Furthermore, there are signals that holography may be operating in this
scenario, although in a much different guise than in AdS/CFT.

In deconstruction, an infrared theory is placed on a one-dimensional lattice.
The link fields that connect adjacent lattice sites provide a Goldstone
realization of an ultraviolet theory in one lower dimension.
The continuum limit of the lattice theory {\it dynamically} generates an
additional spatial direction in the {\it infrared}.
Gravity can be studied within this formalism \cite{harvard, jlm1, jlm2}.
Remarkably, it can be argued that a four-dimensional quantum theory of
gravitation emerges as the infrared limit of coupled $(2+1)$-dimensional
theories of gravity on a lattice \cite{jlm1, jlm2}.
The Bekenstein-Hawking entropy formula (up to the purely numerical factor) is
a {\it universal} statement about the mixing of the UV and IR physics, which
violates the basic principles of a local effective field theory \cite{harvard,
jlm2}.

More explicitly, assuming a local spatial foliation of spacetime, the
Einstein-Hilbert action
\be
S = \frac{1}{G_N} \int \epsilon_{abcd} e^a \wedge e^b \wedge R^{cd},
\ee
expressed in terms of the vierbein and curvature, is classically the
deconstructed version of $N$ copies of three-dimensional general relativity
(a Chern-Simons theory)
coupled to a set of three-dimensional currents \cite{jlm1}.
The parameters of the three-dimensional theory are regarded as fundamental.
The four-dimensional Newton constant is a derived quantity that is determined
by the three-dimensional Newton constant and the lattice spacing $a$:
$G_N =1/M_{Pl}^2= G_3 a$.
Four-dimensional matter fields may also be defined in terms of a
deconstructed three-dimensional theory \cite{nima}.

In a perturbative quantum theory of gravity, the exchange of gravitons ---
local, propagating degrees of freedom --- mediates the dynamical response
of spacetime to the presence of energy and, conversely, the dynamical
response of matter to the geometry of spacetime.
However, in $2+1$ dimensions, gravity is purely topological \cite{witcs}.
There are no local degrees of freedom at all.
To recover the local character of gravitational dynamics in $3+1$ dimensions,
one needs the non-gravitational part of the ultraviolet completion.
Indeed, ``most'' of four-dimensional gravity is reconstructed from the
matter sector (the link fields) of the lattice realization.
These are precisely the three-dimensional currents in our construction
\cite{jlm1}.
The infrared theory organizes this co-dimension one skeleton into the
architecture of spacetime making four-dimensional Lorentz invariance an
emergent property of the continuum limit.

One of the outstanding features of this construction is that it offers a
new viewpoint on the cosmological constant problem \cite{jlm1, jlm2}.
In the deep ultraviolet, there are $N$ essentially independent copies of 
three-dimensional gravity coupled to three-dimensional sources.
These sources induce a conical geometry whose deficit angle prohibits
spinor fields with covariantly constant asymptotics \cite{hen}.
This means that unbroken global supercharges do not exist.
We can have a supersymmetric vacuum without mass degenerate Bose/Fermi
excitations \cite{wit1}.
Three-dimensional supersymmetry therefore implies that the vacuum energy
exactly vanishes at each lattice site.
In the range of intermediate scales, there are $N$ linked copies of
three-dimensional gravity, now coupled to three-dimensional currents.
The geometry is again conical, and the vacuum energy still vanishes.
In the infrared, we recover four-dimensional general relativity with
non-zero cosmological constant.
This is the consequence of a gravitational see-saw, which balances the
Planck mass against the infrared scale $\Delta m$ determined by the
Bose/Fermi mass splitting \cite{jlm2}.

The crucial observation here is that the infrared dynamics ties together
intimately with the physics in the ultraviolet regime.
A tree-level computation indicates that amplitudes involving the
longitudinal components of gravitons de-unitarize at a scale
\cite{harvard, jlm2}
\be
\mu \sim \left(\frac{M_{Pl}^2}{L^5 a^2}\right)^{1/9},
\ee
where $L = Na$ is the lattice size.
By demanding that the theory is truncated above the most massive Kaluza-Klein
states but below the unitarity threshold, we find that the maximum possible
cutoff is of order
\be
\mu_{max} \sim \left(\frac{M_{Pl}^2}{L}\right)^{1/3}.
\label{eq:cutoff}
\ee
This exemplifies the phenomenon of UV/IR mixing:
the ultraviolet cutoff is defined in terms of purely infrared quantities,
namely the size of the extra spatial dimension that arises from
deconstruction and the four-dimensional Planck mass.\footnote{UV/IR mixing
also signifies non-locality in the effective action for the Kaluza-Klein
modes \cite{harvard}.}

The scale $\mu_{max}$ has an important holographic interpretation.
We can compute the entropy using the thermodynamic relation
${\cal S} \sim V T^3$,
where the volume $V \sim AL$ and the temperature $T \sim \Lambda_{max}$,
the ultraviolet cutoff.
Taking $\Lambda_{max} \sim M_{Pl}$ yields the standard wrong result,
but if instead, we use the expression from \eref{cutoff}, we find that 
\be
{\cal S} \sim AL \mu_{max}^3 \sim \frac{NA}{G_3 L} \sim \frac{A}{G_N},
\ee
which is nothing but the holographic bound on the number of degrees of
freedom in the ultraviolet theory, as it {\it must} be if deconstruction
is expected to provide an ultraviolet definition of four-dimensional gravity.
We conclude that unitarity plus diffeomorphism invariance are sufficient to
imply holography.
The argument generalizes to an arbitrary number of dimensions.

The infrared theory ({\em i.e.}, the {\it four}-dimensional continuum limit
of the lattice theory) lies in the region where the {\it three}-dimensional
interaction strength is strongly coupled.
The cosmological constant problem is to explain why the vacuum energy is
small but non-vanishing at long distances in this region of strong coupling.

There are two natural mass scales in the infrared.
Each of these arise from the dimensionful parameters in the ultraviolet,
the lattice spacing $a$ and the Newton constant $G_3$.
One scale is simply the four-dimensional Planck mass, $M_{Pl}$, which sets
the strength of the gravitational interaction.
Since $G_N = G_3 a$, the three-dimensional scale is much higher than the
effective four-dimensional gravitational scale as we approach the continuum.
Thus, $M_{Pl}$ is indeed an infrared scale from the three-dimensional
point of view.
A second low-energy scale, $\Delta m$, is defined by the mass
difference between Bose and Fermi excitations in the three-dimensional
theory. 
Given these two scales and the requirement that the vacuum energy vanishes
in the limit where the mass splitting between bosonic and fermionic degrees
of freedom goes to zero, we can associate a single scale $\omega$ with
$M_{Pl}$ and $\Delta m$.
This scale serves as a cutoff in the computation of the four-dimensional
vacuum energy.
Dimensional analysis informs us that
\be
\omega \sim \frac{(\Delta m)^2}{M_{Pl}}.
\ee 
Since $\omega$ represents the ultraviolet scale in the computation of the
vacuum energy density and is determined by infrared quantities $M_{Pl}$
and $\Delta m$, this relation is also a manifestation of the UV/IR
correspondence.

When evaluating vacuum diagrams in order to estimate the upper bound on
the vacuum energy in the infrared, we use $\omega$ as the only effective
cutoff in the theory.
The na\"{\i}ve expression\footnote{Of course, one should make a careful study
of radiative corrections as well, even though these cannot be disastrous
if we remember that the vacuum energy is {\it zero}, by deconstruction, 
down to a very low energy scale.} 
for the vacuum energy is bounded by $\omega^4$, or
\be
\Lambda \sim M_{Pl}^4 \left(\frac{\Delta m}{M_{Pl}}\right)^8.
\ee
Therefore, the observed bound on the vacuum energy density can be realized
by a large separation between the mass splitting and the Planck scale.

This argument relies upon a few basic assumptions:  dimensional analysis,
the UV/IR relation we have discussed previously, three-dimensional
supersymmetry, and the notion that the deconstruction of Witten's argument
for the vanishing of the cosmological constant in $2+1$ dimensions implies
zero vacuum energy at a very low scale set by $\Delta m$.
The limit $\Delta m \to 0$, in which the four-dimensional cosmological
constant vanishes, corresponds to the restoration of the mass degeneracy
in three dimensions.
This observation is consistent with the principle that vanishing dimensionful
parameters correspond to enhanced symmetries.

Deconstruction offers a new way of dealing with the famously difficult
ultraviolet problems of four-dimensional gravity.
In this approach, four-dimensional Lorentz invariance is an emergent
symmetry.
Deconstruction also leads to a new perspective on the holographic principle
as well as on one of the outstanding puzzles of fundamental physics,
the cosmological constant problem.
The numerical smallness of the cosmological constant is implied by a unique
combination of holography and supersymmetry.
Given that the total energy density of the universe today is apparently
dominated by the energy in the vacuum and the critical r\^ole supersymmetry
has played in the effort to understand the stability of the vacuum, we
expect that the deconstruction of four-dimensional gravity will in the
future provide many additional insights about Nature.

\section*{Acknowledgments}
\noindent 
This work is supported in part by the U.S.\ Department of Energy under
contracts DE-FG02-91ER40677 (RGL).


\begin{thebibliography}{99}

\bibitem{holo}
G.~'t Hooft,
arXiv:gr-qc/9310026;
L.~Susskind,
J.\ Math.\ Phys.\ {\bf 36}, 6377 (1995)
[arXiv:hep-th/9409089].

\bibitem{bhent}
J.~D.~Bekenstein,
Phys.\ Rev.\ D {\bf 7}, 2333 (1973);
J.~M.~Bardeen, B.~Carter and S.~W.~Hawking,
Commun.\ Math.\ Phys.\  {\bf 31}, 161 (1973);
S.~W.~Hawking,
Commun.\ Math.\ Phys.\  {\bf 43}, 199 (1975).

\bibitem{ads}
J.~Maldacena,
Adv.\ Theor.\ Math.\ Phys.\  {\bf 2}, 231 (1998)
[arXiv:hep-th/9711200];
S.~S.~Gubser, I.~R.~Klebanov, A.~M.~Polyakov,
Phys.\ Lett.\ {\bf B428} (1998) 105 [arXiv:hep-th/9802109];
E.~Witten, 
Adv.\ Theor.\ Math.\ Phys. {\bf2}(1998) 253
[arXiv:hep-th/9802150].

\bibitem{WMAP}
C.~L.~Bennett {\it et al.},
arXiv:astro-ph/0302207.

\bibitem{SN}
A.~G.~Riess {\it et al.}  [Supernova Search Team Collaboration],
Astron.\ J.\  {\bf 116}, 1009 (1998)
[arXiv:astro-ph/9805201];
S.~Perlmutter {\it et al.}  [Supernova Cosmology Project Collaboration],
Astrophys.\ J.\  {\bf 517}, 565 (1999)
[arXiv:astro-ph/9812133].

\bibitem{dscft}
C.~M.~Hull,
JHEP {\bf 9807}, 021 (1998)
[arXiv:hep-th/9806146];
V.~Balasubramanian, P.~Horava and D.~Minic,
JHEP {\bf 0105}, 043 (2001)
[arXiv:hep-th/0103171];
E.~Witten,
hep-th/0106109;
A.~Strominger,
JHEP {\bf 0110}, 034 (2001)
[arXiv:hep-th/0106113];
R.~Bousso, A.~Maloney and A.~Strominger,
Phys.\ Rev.\ D {\bf 65}, 104039 (2002)
[arXiv:hep-th/0112218];
V.~Balasubramanian, J.~de Boer and D.~Minic,
Phys.\ Rev.\ D {\bf 65}, 123508 (2002)
[arXiv:hep-th/0110108];
V.~Balasubramanian, J.~de Boer and D.~Minic,
Annals Phys.\  {\bf 303}, 59 (2003)
[arXiv:hep-th/0207245].

\bibitem{nima}
N.~Arkani-Hamed, A.~G.~Cohen and H.~Georgi,
Phys.\ Rev.\ Lett.\  {\bf 86}, 4757 (2001)
[arXiv:hep-th/0104005].

\bibitem{wit1}
E.~Witten,
Int.\ J.\ Mod.\ Phys.\ A {\bf 10}, 1247 (1995)
[arXiv:hep-th/9409111].
See also
E.~Witten,
Mod.\ Phys.\ Lett.\ A {\bf 10}, 2153 (1995)
[arXiv:hep-th/9506101].

\bibitem{harvard}
N.~Arkani-Hamed, H.~Georgi and M.~D.~Schwartz,
arXiv:hep-th/0210184;
N.~Arkani-Hamed and M.~D.~Schwartz
[arXiv:hep-th/0302110].

\bibitem{jlm1}
V.~Jejjala, R.~G.~Leigh and D.~Minic,
Phys.\ Lett.\ B {\bf 556}, 71 (2003)
[arXiv:hep-th/0212057].

\bibitem{jlm2}
V.~Jejjala, R.~G.~Leigh and D.~Minic,
arXiv:hep-th/0302230.

\bibitem{witcs}
E.~Witten,
Nucl.\ Phys.\ B {\bf 311}, 46 (1988);
J.~H.~Horne and E.~Witten,
Phys.\ Rev.\ Lett.\  {\bf 62}, 501 (1989).
See also
A.~Achucarro and P.~K.~Townsend,
Phys.\ Lett.\ B {\bf 180}, 89 (1986).

\bibitem{hen}
M.~Henneaux,
Phys.\ Rev.\ D {\bf 29}, 2766 (1984).




\end{thebibliography}
\end{document}